\documentclass[preprint,12pt]{elsarticle}

\journal{Physics of Life Reviews}

\begin{document}

\begin{frontmatter}

\title{From Physics to Sentience: Deciphering the Semantics of the Free-Energy Principle and Evaluating its Claims\\
Comment on "Path integrals, particular kinds, and strange things" by Karl Friston \textit{et al.}}
\author[1]{Zahra Sheikhbahaee\fnref{*}}
\author[2]{Adam Safron\fnref{*}}
\author[3]{Casper Hesp}
\author[1,4]{Guillaume Dumas}
\affiliation[1]{CHU Sainte-Justine Research Center, Department of Psychiatry, University of Montreal}
\affiliation[2]{Center for Psychedelic & Consciousness Research, John Hopkins University}
\affiliation[3]{Department for Developmental Psychology, University of Amsterdam}
\affiliation[4]{Mila -- Quebec AI Institute}
\date{October 2023}

\fntext[*]{These authors contributed equally to this work}

\end{frontmatter}

The Free-Energy Principle (FEP) \cite{Friston23,friston2007free, friston2019free} has been adopted in a variety of ambitious proposals that aim to characterize all adaptive, sentient, and cognitive systems within a unifying framework. Judging by the amount of attention it has received from the scientific community, the FEP has gained significant traction in these pursuits. The current target article represents an important iteration of this research paradigm in formally describing emergent dynamics rather than merely (quasi-)steady states. This affords more in-depth considerations of the spatio-temporal complexities of cross-scale causality – as we have encouraged and built towards in previous publications (e.g., \cite{Bruineberg2018,hesp2022b,hesp2022,Fristonetal2021,Kastel&Hesp2021,kelso2013outline}). In this spirit of constructive feedback, we submit a few technical comments on some of the matters that appear to require further attention, in order to improve the clarity, rigour, and applicability of this framework. 

Certain idiosyncratic definitions and derivative statements in this paper and prior works have tended to confuse scientists who attempt to assess the validity of such formulations in depth. As the authors aimed to formulate a robust Bayesian mechanics by borrowing from physics, particular care needs to be taken concerning physics terminology.

In the target article, much of the authors' argument relies on what they define as ``generalized coordinates", effectively a vector containing higher-order time derivatives of motion of a (sub)system, which can be used to approximate its local dynamics under certain limiting assumptions, on some (unspecified) time scale. While the authors' usage appears to be internally consistent, it may generate confusion with respect to a particular analytical method that physicists hold dear. When grappling with complex dynamical systems, many hours of work can be saved by applying a state-space reduction and exploiting the coordinate-agnostic flexibility of Lagrangian mechanics, where ``generalized coordinates" and ``generalized velocities" refer to $q$ and $\dot{q}$, respectively, in the Lagrange equation of the second kind:  $\frac{d}{dt} (\frac{\delta L}{\delta \dot{q}_j}) = \frac{\delta L}{\delta q_j}$.

In that context, carefully choosing a set of generalized coordinates is a highly context-specific and technical task in itself that can simplify derivations of physically meaningful constraints on a system's dynamics (e.g., a textbook example: shifting to angular coordinates to effectively model a pendulum; see \cite{goldstein2002classical}). If the authors intend to engage the physics community effectively and avoid unnecessary confusion for novel readers, it would seem expedient to choose a different label, or else clearly explain differences from common (physics-textbook) linguistic use.

Such an apparently innocuous mixing of terms can engender further misunderstandings, leading to breakdowns in communication between fields. For example, consider the following quote from the target article: \textit{``In this setting, the Lagrangian plays the role of an action, where paths of least action minimize the Lagrangian of generalized states." (p.36)} This quote would make most physicists frown because the action is commonly defined as the path integral of the Lagrangian – as acknowledged by the authors (see also \cite{landau1976mechanics}). Within the confines of the target article, however, ``generalised states" are defined such that they implicitly characterize local paths, which means that the Lagrangian of all derivatives can be seen as specifying an (implicit) action of sorts (valid only to the extent that a Taylor series approximation holds).

More generally speaking, many claims throughout the article contain tacit assumptions, often leaving the reader to reverse engineer what additional details the authors might have had in mind that could help constrain their claims. For example, consider this sentence:
\textit{``We make no assumptions about the solutions of this equation, other than the flow operator does not change over some relevant time interval."}(p.38). Since the formulation itself consists of postulates and approximations with unclear boundary conditions, it becomes difficult (or impossible) to evaluate individual claims. In their pursuit of generality, most concepts have been left either unconstrained or defined in a self-referential manner. In extension, the same holds for solutions, as exemplified above: the authors assumed that solutions have flow operators that do not change on relevant time intervals without specifying what degrees of change might be considered significant and what kinds of time intervals could be ``relevant". One might wonder under which conditions a Taylor series will (still) provide satisfactory approximations when considering systems that display cross-scale itinerancy. There are likely dynamical regimes where such polynomial approximations must be extended with periodic components, but currently, the blanket response would be to simply postulate ``relevant" time scales at which the Taylor series is enough.

The need for precise semantics becomes critical when the FEP claims delve into complicated topics such as agency and sentience. The words autonomy and agency are used interchangeably, but there is a significant difference between systems driven by meaningful internal dynamics~\cite{Rovelli18}, and systems that are capable of planning in order to achieve explicitly represented goals; thus, it remains important to recognize the difference between teleology vs. 'mere' telonomy \cite{Humphrey2023,Safron21Embodied,Deacon11}.

In the article, the authors invoke notions of "strange particles" and the circular causation involved in their beliefs about their self-generated actions. This handling of "self-evidencing"~\cite{Hohwy2016} may not only help speak to the ``strange loops" characteristic of some complex adaptive systems, but also to the deeper issues of life-mind continuity \cite{thompson2010mind}, and the recursive nature of consciousness \cite{hofstadter1979godel,thompson2001radical}.

The authors also allude to the concept of consciousness (using quotes from Heisenberg and David Bohm), but never clarify how it relates to their definition of ``sentience". Connecting the behaviour of strange particles with 'sentience' might be premature, depending on the semantic scope of this concept. While self-evidencing systems can be considered to be sentient in that they are governed by information given valence by reference to self-persistence, it is important not to conflate this with consciousness, lest we risk being so inclusive with this concept as to compromise the utility of considering the conscious properties of a system. If conscious brains are clearly a self-referential system, not all self-referential systems are necessarily sentient or conscious.

Conscious experience likely requires complex network architectures capable of supporting ``integrated world modeling" ~\cite{Safron22IWMT,Safron2020}. Such functional properties may potentially be achieved in a variety of ways in different kinds of "particles," such as through information bottlenecks emerging from complex attractor dynamics~\cite{Volzhenin22PNAS}. However, to reliably ground such inferences about the potential conscious properties of a system, it is required that we both deeply engage with models of consciousness for systems for which we know there are subjective experiences, as well as precisely characterize the phenomenology (in the sense used by physicists) of systems that may realize such qualitative aspects by other means."

To associate the functionalities (and perhaps phenomenality) of "sentience" with particles may generate further confusion by both implying the kinds of panpsychism the authors explicitly renounce, as well as potentially suggesting a kind of illegitimate eliminative reductionism concerning the values of agents with conscious experience (e.g. the "as if" language used throughout the paper for the ontological status of particles with goals).

In conclusion, this FEP-based formalism presented by Friston and colleagues offers an intriguing perspective on life and cognition, but many conceptual nuances still need to be addressed. This painstaking process could eventually allow the authors to do justice to the complex systems they want to describe and the global and interdisciplinary community of academics they aim to engage.

\section*{Acknowledgement}
Zahra Sheikhbahaee and Adam Safron were supported by the Survival and Flourishing Fund (SFF-2023-H1) through the Institute for Advances Consciousness Studies (IACS). Guillaume Dumas was supported by the Institute for Data Valorization, Montreal (IVADO; CF00137433), the Fonds de recherche du Qu\'{e}bec (FRQ; 285289), the Natural Sciences and Engineering Research Council of Canada (NSERC; DGECR-2023-00089), and the Azrieli Global Scholars Fellowship from the Canadian Institute for Advanced Research (CIFAR) in the Brain, Mind, \& Consciousness program. Additionally, the authors gratefully acknowledge the helpful feedback provided by colleagues, particularly Vincent Taschereau-Dumouchel. The author has no conflicts of interest to disclose.

\newpage
\bibliographystyle{elsarticle-num}
\bibliography{main}

\end{document}